\documentclass[amsfonts,amssymb,amsmath,twocolumn, showkeys, nofootinbib, superscriptaddress]{revtex4-1}
\usepackage{amsthm}
\usepackage{mathtools}
\usepackage{xcolor}
\usepackage{graphicx}
\usepackage[left=23mm,right=13mm,top=35mm,columnsep=15pt]{geometry} 
\usepackage{adjustbox}
\usepackage{placeins}
\usepackage[T1]{fontenc}
\usepackage{lipsum}
\usepackage{csquotes}
\usepackage{tikz}
\usepackage{quantikz}
\usepackage{xcolor}
\usepackage[pdftex, pdftitle={Article}, pdfauthor={Author}]{hyperref} 
\usepackage{ulem}
\definecolor{nice-blue}{HTML}{1F77B4}
\definecolor{nice-green}{HTML}{2CA02C}
\definecolor{nice-red}{HTML}{D62728}
\definecolor{nice-orange}{HTML}{FF7F0E}

\begin{document}

\title{Automatic post-selection by ancillae thermalization}

\author{L. Wright}
\email{lewis.wright@kcl.ac.uk}
\affiliation{Department of Physics, King's
  College London, Strand, London WC2R 2LS, United Kingdom}
  
\author{F. Barratt}
\affiliation{Department of Mathematics, King's
  College London, Strand, London WC2R 2LS, United Kingdom}

\author{J. Dborin}
\affiliation{London Centre for Nanotechnology,
  University College London, Gordon St., London, WC1H 0AH, United
  Kingdom}
  
\author{G. H. Booth}
\affiliation{Department of Physics, King's
  College London, Strand, London WC2R 2LS, United Kingdom}
 
\author{A. G. Green}
\affiliation{London Centre for Nanotechnology,
  University College London, Gordon St., London, WC1H 0AH, United
  Kingdom}
 
\date{\today} % Leave empty to omit a date

\begin{abstract}
Tasks such as classification of data and determining the groundstate of a Hamiltonian cannot be carried out through purely unitary quantum evolution. Instead, the inherent non-unitarity of the measurement process must be harnessed. Post-selection and its extensions provide a way to do this. However they make inefficient use of time resources --- a typical computation might require $O(2^m)$ measurements over $m$ qubits to reach a desired accuracy and cannot be done intermittently on current (superconducting-based) NISQ devices.
We propose a method inspired by thermalization that harnesses insensitivity to the details of the bath.
We find a greater robustness to gate noise by coupling to this bath, with a similar cost in time and more qubits compared to alternate methods for inducing non-linearity such as fixed-point quantum search for oblivious amplitude amplification.
Post-selection on $m$ ancillae qubits is replaced with tracing out $O\left(\log\epsilon / \log(1-p)\right)$ (where p is the probability of a successful measurement) to attain the same accuracy as the post-selection circuit.
We demonstrate this scheme on the quantum perceptron, quantum gearbox and phase estimation algorithm. 
This method is particularly advantageous on current quantum computers involving superconducting circuits.  
\end{abstract}

\maketitle

\section{Introduction}
%%%%%%%%%%%
 \label{sec:Introduction}
Algorithms for classification of data, optimising the energy to find the groundstate properties of a Hamiltonian (and indeed optimising classifiers for a given data set) require the use of non-linear operations that cannot be achieved solely through unitary quantum evolution.
When carrying out these tasks on a quantum computer we must use the non-unitarity of the measurement process. There are several ways in which to do this depending upon the relative abundance of resources, quantified by measures such as the number of qubits, coherence times and gate fidelities. Post-selection and the related repeat-until-success algorithms are  popular choices.

However post-selection makes inefficient use of time resources --- a typical computation requires $O(2^m)$ measurements over $m$ qubits to reach a desired accuracy. Nevertheless, it is a frequently used tool in atomic contexts where coherence times are long and manipulation timescales short, so that time is not the limiting resource. For superconducting circuits where coherence times are much shorter and where measurements of a subset of qubits is not possible while maintaining the coherent evolution of the remainder, this is  more problematic. 

Curiously, post-selection in fact uses classical non-linearity --- through a yes or no decision based upon a measurement on ancillae qubits. This suggests how a more time-efficient scheme might be developed. Fundamentally, the non-linearity of the classical world is induced by observation of only a portion of a larger quantum world. It is possible then to replace post-selection with a scheme where explicit measurement of ancillae qubits is not required, {\it i.e.} where they are traced out or simply ignored.

Eigenstate thermalization gives a clue as to how this can be achieved. Coupling a small system at high temperatures to a large, low temperature bath allows us to cool the small system. Eigenstate thermalization extends this notion to closed quantum systems. Coupling a large number of ancillae qubits in a low entropy state ({\it e.g.} $|00000... \rangle $) to a small system and evolving the total system under some unitary evolution allows entropy to flow from the small system of interest to the ancillae. 
\begin{figure}[h!]
%\vspace{-3cm}
\hspace{-2.85cm}\raisebox{50pt}{\textbf{a)}}
\includegraphics[width = 5.25cm]{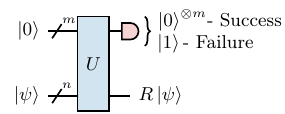}
\\
\vspace{0.1cm}
\hspace{-0.6cm}
\raisebox{110pt}{\textbf{b)}}
\includegraphics[width = 7.5cm]{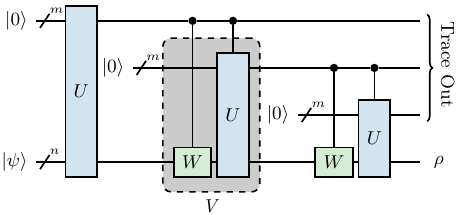}
\caption{\textbf{Comparison between post-selection and ancillae thermalization.}
\textbf{a)} Post-selection circuit acting on $n$-qubit target state $\ket{\psi}$ and $m$ ancillae.
 Applying $U$ and measuring $\ket{0}^{\otimes m}$ on the ancillae guarantees a successful transformation $R\ket{\psi}$ of the target qubits. Measuring $\ket{1}$ on any ancilla implies a failure and the procedure is repeated.
 \textbf{b)} Ancillae thermalization circuit equivalent to
 $O(N)$ attempts at applying $\ket{\text{R}\psi}$ with $N=2$. Note that each attempt of post-selection is exponential in measurements.
 Ancillae thermalization is a modification of post-selection. Measurements are replaced with gates and new ancillae entangle with the incorrectly transformed parts of the target qubits' wavefunction. All ancillae are then traced out. The unitary $V$ acts on $\ket{\psi}$ and $m$ ancillae per iteration. It is factorised as $V = UW$ where $W$ is a reset gate discussed in Appendix~\ref{appendix:scramb and nor}. The circuit produces a finalised mixed target state $\rho$ that has an overlap with desired state $R\ket{\psi}$ up to error $\epsilon$, i.e. $\text{Tr}(\rho \ket{\psi'}\bra{\psi'}) = 1-\epsilon$ where $\ket{\psi'} = R \ket{\psi}$.}
\label{fig:General post-selection vs tracing out circuit}
\end{figure}
\section{Results}
%%%%%%%%%
Inspired by this, we replace the classical yes or no non-linearity of post-selection with a non-linearity attained by tracing out ancillae.
In the following, we demonstrate the application of ancillae thermalization to the quantum perception, quantum gearbox and a groundstate preparation algorithm.
Robustness to noise is demonstrated for the quantum gearbox in simulation for two different noise models and on the `ibmq\_virgo' quantum device. 
\\
Our scheme is a type of amplitude amplification- a generalisation of Grover search \cite{grover1996fast} introduced by Brassard \textit{et. al} \cite{brassard2002quantum}.
Amplitude amplification comes in various flavours depending upon whether the state of the target qubits or probability of success at each iteration is known or not.
For a direct comparison with our method, we focus upon an implementation that does not require the knowledge of either and decreases the error monotonically,
$\frac{\pi}{3}$ fixed-point search \cite{grover2005fixed}.
Used in the context of amplitude amplification, this algorithm which we call $\frac{\pi}{3}$ fixed-point oblivious amplitude amplification (FP OAA) is equivalent to the optimal `Fixed-point quantum search' from Yoder \textit{et. al} in the regime of large initial success probabilities \cite{yoder2014fixed}.
In this regime we find ancillae thermalization more robust to the effects of gate errors, with a similar cost in time and more qubits.
In addition, the different structure of our method allows implementation of a wider class of transformations whilst retaining no knowledge of the target state, including non-unitary transformations, the latter of which OAA algorithms alone do not allow \cite{berry2014exponential}.
We demonstrate this using a non-optimal procedure for groundstate preparation shown in Fig.\ref{fig:Phase estimation}.
Table~\ref{tab:General post-selection vs tracing out circuit} and Fig.~\ref{fig:at vs oaa noise} show the trade-offs in resources between the various approaches for unitary transformations and robustness to gate noise respectively, in the large success probability regime.

\subsection{From post-selection to ancillae thermalization} \label{subsec: Post Select & Tracing Out}
%%%%%%%%%%%%%%%%%%%%%%%%%%%%%%%%%%%%%%%%%%%%%%%%%%%%%%%%
Fig.~\ref{fig:General post-selection vs tracing out circuit}a) demonstrates post-selection used in a repeat until success circuit. A unitary $U$ applies a desired operation $R$ to an input state $\ket{\psi}$, conditioned upon the outcome of a measurement of $m$ ancillary qubits. For example the unitary,
\begin{equation}
 U \ket{0}^{\otimes m} \ket{\psi} 
 = 
 \sqrt{p_0} \ket{0}^{\otimes m} R\ket{\psi} + \sum_{k=1}^{2^m - 1} \sqrt{p_k} \ket{k} E_{k}\ket{\psi},
\end{equation}
achieves the rotation $R$ on $\ket{\psi}$ and the state $\ket{0}^{\otimes m}$ on the ancillary qubits with probability $p_0$. The states $E_{k}\ket{\psi}$ corresponds to incorrect transformations of the target qubits.
If the procedure fails, all qubits are reset and the process is repeated. The probability of failure after $N$ iterations of the algorithm is $\epsilon \sim (1-p_0)^N$; the transformation $R$ is implemented {\it exactly} with a finite probability. 

Ancillae thermalization and more generally amplitude amplification take a different philosophy. The output from $U$ on the target qubits is interpreted as a superposition of correctly transformed ($R\ket{\psi}$) and incorrectly transformed ($E\ket{\psi}$) states. Tracing out the ancillary qubits prepares the target qubits in a mixed state $\rho$ that is {\it approximately correct}.
The density matrix $\rho$ has an overlap $\bra{\psi} R^\dagger \rho R \ket{\psi}=1-\epsilon$ with $R\ket{\psi}$.
For ancillae thermalization, this fidelity with the target state is obtained with an exponential reduction in measurements for an increased cost in ancillary qubits. 
This is achieved by iteratively entangling fresh ancillary qubits with the target qubits via unitary $V=UW$, where $W$ is a reset gate that transforms $E_k \ket{\psi} \rightarrow \ket{\psi}$ for all $k$.
The ancillae conditionally entangle with the incorrectly transformed parts of the system's wavefunction through control gates on the previous ancilla at each iteration.
The circuit to achieve this is shown in Fig.~\ref{fig:General post-selection vs tracing out circuit}b).
The overlap between the target qubits and $R\ket{\psi}$ increases exponentially with the number of iterations applied (see Appendix~\ref{appendix:scramb and nor} for full details). 

\begin{table*}[ht]
\hfill
\begin{tabular}[t]{l|c|c|c}
Method\ \ \ \ \  & Measurements & Qubits & Gates\\
 \hline
  Post-Selection & $\displaystyle O\left(p_0^{-1}\right)$ & $n + m$ & $\displaystyle Q(U_{n+m})$ \\
 \begin{tabular}[c]{@{}l@{}}$\frac{\pi}{3}$ FP OAA\end{tabular}  &  $\displaystyle 0$  & $\displaystyle O\left(n + m\right)$ & 
 $\displaystyle O\left(\frac{\log\epsilon}{\log(1-p_0)} [m + Q(U_{n+m})]\right)$\\
 \begin{tabular}[c]{@{}l@{}}Ancillae\\Thermalization\end{tabular} &  $\displaystyle 0$ & $\displaystyle O\left(n+ \frac{\log\epsilon}{\log(1-p_0)} m\right)$& 
 $\displaystyle O\left(\frac{\log\epsilon}{\log(1-p_0)} [m + Q(U_{n+m}) + Q(W_n)]\right)$ 
\end{tabular}\hfill \qquad\qquad\qquad

\caption{\textbf{Comparison of Computational Resources for Unitary Transformations.} Computational resources in the of post-selection, ancillae thermalization and $\pi/3$ fixed-point oblivious amplitude amplification (FP OAA) \cite{grover2005fixed} for guaranteeing unitary transformations in the large initial success probability regime. We define $Q(O_{N})$ as the number of operations required to implement the N-qubit gate, $O$, and $p_0$ as the initial probability of a successful measurement.
In the asymptotic limit the number of operations required to implement controls on $U_{n+m}$ is constant. Additionally, $Q(W_n) \sim O(1)$ and can be ignored in practice.
Although ancillae thermalization has more operations, we observe lower susceptibility to gate errors.
We suspect this is due to the exponentially fewer operations acting on each ancilla in ancillae thermalization, exposing them to less gate noise. Justification for these values can be found in Appendix~\ref{Appendix:Resource Scaling} whilst a comparison of noise robustness can be found in section~\ref{appendix:noise}.
}
\label{tab:General post-selection vs tracing out circuit} 
\end{table*}
%number of operations to implement controls on target qubit specific gates.
\subsection{Quantum Perceptron}
%%%%%%%%%%%%%%%%%
%
\vspace{1cm}\begin{figure}[h!]
\hspace{-1.9cm}\raisebox{30pt}{\textbf{(a)}}
 \begin{quantikz}[row sep=0.2cm, column sep=0.2cm]
 \lstick{\ket{0}} & \qw&\gate{R_y(2\theta)}\gategroup[2,steps=3,style={dashed,rounded corners,fill=nice-blue!20, inner xsep=2pt},
 background,label style={label position=below,anchor=north,yshift=-0.2cm}]{\textit{U}} & \ctrl{1} & \gate{R_y^{\dagger}(2\theta)} & \qw&\meter{}  \\
 \lstick{\ket{\psi}} &\qw& \qw& \gate{-iY}& \qw  & \qw &\qw
 \end{quantikz}
 \\
 \hspace{-0.6cm}\raisebox{65pt}{\textbf{(b)}}
\includegraphics[width = 7.5cm]{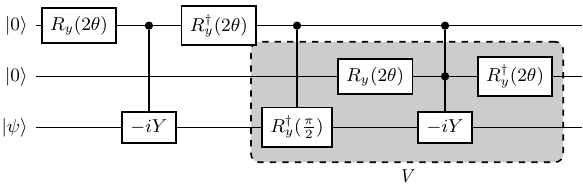}
\\
\hspace{-0.6cm}\raisebox{160pt}{\textbf{(c)}}
\includegraphics[width = 7.5cm]{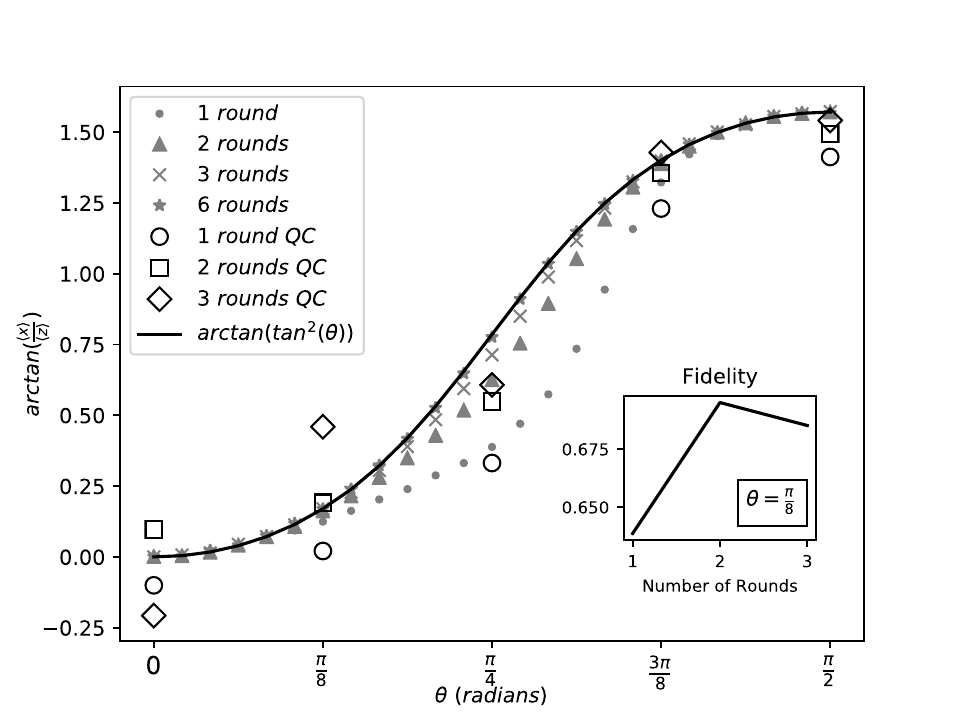}
 \caption{\textbf{Quantum Perceptron/non-linear activation function $q(\theta)$.}
 \textbf{a)} Post-select circuit for implementing angle $q(\theta)$ in the quantum perceptron, acting on an ancilla and the target qubit. A successful transformation of $\exp{\text{(-}iq(\theta)Y}\text{)}|\psi\rangle$ corresponds to measuring $\ket{0}$ on the ancilla with probability $p(\theta)\text{=}\cos^4(\theta)\text{+}\sin^4(\theta)$. Upon failure, when $\ket{1}$ is measured on the ancilla, the target qubit is guaranteed to transform as $\exp{\text{(-}i\pi/4\text{)}}|\psi\rangle$. As a result, the target qubit can be reset by rotation $R_y(\pi/2)$ and the circuit is repeated.
 \textbf{b)} Ancillae thermalization circuit for an equivalent $O(N)$ attempts of post-selection, with $N=2$ applications of the circuit in a). The reset and second instance, which acts on a new ancilla, are both conditioned by the state of the first ancilla. 
\textbf{c)} Angle $q(\theta)$ obtained by ancillae thermalization for different number of iterations and $\theta$.
The values were obtained from IBMQ's `qasm-simulator' (symbols) (no noise) and `ibmq\_ourense' quantum computer (rings).
The subfigure shows the fidelity from `ibmq\_ourense' between the finalised target qubits and groundstate for different numbers of iterations.}
\label{fig:Quantum Perceptron}
\end{figure}
The quantum perceptron\cite{cao2017quantum} is the first explicit example of a quantum circuit fulfilling the requirements for a meaningful quantum neural network.
It was introduced by Schuld et. al \cite{schuld2014quest}. It is able to simulate a classical perceptron whilst taking advantage of quantum properties such as processing input data as a superposition.
In general quantum neural networks struggle to construct a nonlinear activation function due to their linear dynamics. The quantum perceptron uses a post-select circuit shown in Fig.~\ref{fig:Quantum Perceptron}a) to achieve this non-linearity. This circuit implements the transformation $\ket{\psi} \rightarrow \exp(-iq(\theta)Y)\ket{\psi}$ onto a target qubit with probability $p(\theta) \sim O\left(1/2\right)$. The angle of rotation $q(\theta) = \arctan(\tan^2(\theta))$ is sigmoidal in shape and can be used to capture the non-linear properties found in classical neural networks in a quantum setting.

Ancillae thermalization removes the need to post-select in order implement the quantum perceptron. 
The circuit shown in Fig.~\ref{fig:Quantum Perceptron}b) achieves the same level of accuracy as $O(N)$, with $N=2$ attempts of the post-selection circuit. To achieve a total overlap with the desired state $\exp(-iq(\theta)Y)\ket{\psi}$ within additive error $\epsilon$, the process of applying $V$ to fresh ancillae and the target qubit must be repeated $O(\log(1/\epsilon))$ times.
This achieves a fidelity between the finalised target qubit and the desired state given by,
\begin{eqnarray}
& &    \bra{\psi}e^{iq(\theta)Y}  \rho \;  e^{-iq(\theta)Y} \ket{\psi} 
\nonumber\\
 &=& 1 - \delta^{\frac{\log(1/\epsilon)}{\log(1/\delta)} + 1}\left(1 - |\langle \psi |e^{-iq(\theta)Y}\psi\rangle|^2 \right),
\label{eq:Fidelity Quantum Perceptron}
\end{eqnarray}
where $\delta=1-p(\theta)$ and $\epsilon$ has been re-scaled.
The fidelity increases exponentially with the number of iterations.
\\
\\
Results of applying ancillae thermalization to the quantum perceptron obtained from IBMQ's `qasm-simulator` i.e. a simulator with no noise and `ibmq-oursense' quantum machine are shown in Fig.~\ref{fig:Quantum Perceptron}c).
As in other applications of NISQ devices, there is an optimum circuit depth that balances theoretical advantages of deeper circuits with the effects of noise. The quantum perceptron displays an increase in accuracy with increasing iterations up to a threshold where further operations increase exposure to finite gate fidelity leading to a decrease  in accuracy.  This point is emphasised in the sub-figure which shows a lower fidelity for a higher number of iterations. 

\subsection{Phase Estimation}\label{sec:Phase Estimation}
%%%%%%%%%%%%%%%%%%%%%
%
\begin{figure*}[ht]
\hspace{-1.0cm}
\begin{minipage}[b]{0.45\linewidth}
\centering
\hspace{-8cm}\raisebox{0pt}{\textbf{(a)}}
 \\
\begin{quantikz}[row sep=0.4cm, column sep=0.4cm]
        \lstick[wires=3, label style = {rotate = 90, xshift = 1.1cm,yshift = 0.5cm}]{\text{m ancillae}\\ \text{precision qubits}} & \lstick{\ket{0}} &\qw & \gate{H}\gategroup[4,steps=5,style={dashed,
    rounded corners,fill=nice-blue!20, inner xsep=2pt},
    background,label style={label position=below,anchor=
    north,yshift=-0.2cm}]{\textit{U}} & \qw&  \ \ldots\ \qw & \ctrl{3} & \gate[wires = 3, nwires = 2]{\text{QFT$^{\dagger}$}} & \meter{} \\
         &&&\lstick[label style = {xshift = 0.25cm}]{$\vdots$}&&&&&\lstick[label style = {xshift = 0.25cm}]{$\vdots$}\\
        &\lstick{\ket{0}} &\qw  &\gate{H}  & \ctrl{1} & \ \ldots\ \qw &\qw & \ &\meter{} \\
         &  \lstick{\ket{\psi}}&\qw\qwbundle{n} &\qw  & \gate{\textit{A\,$^{2^0}$}} & \ \ldots\ \qw & \gate{\textit{A\,$^{2^{m-1}}$}} & \qw &\qw
\end{quantikz}
\\
\hspace{-7.75cm}\raisebox{40pt}{\textbf{(b)}}
\\
\vspace{-1.1cm}\includegraphics[width = 9.3cm]{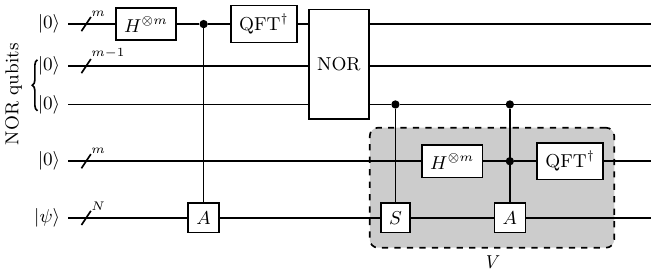}
\end{minipage}
\hspace{2.0cm}
\begin{minipage}[c]{0.45\linewidth}
\hspace{-7.0cm}\raisebox{0pt}{\textbf{(c)}}
\\
\hspace*{0cm}
\vspace{10cm}\includegraphics[width = 9.0cm]{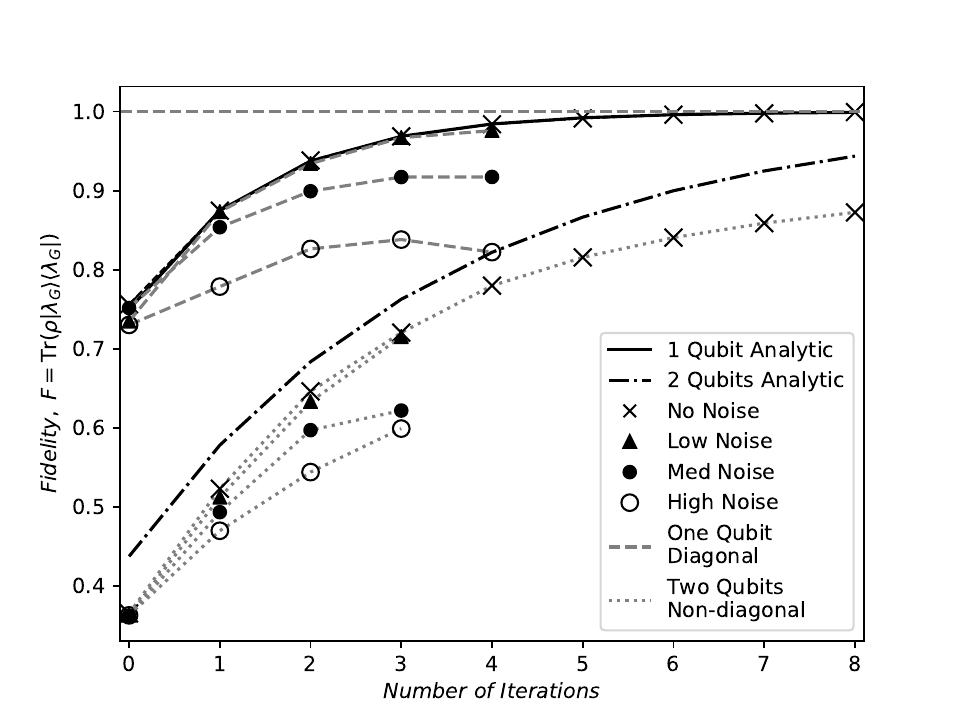}
\end{minipage}
\vspace{-8.7cm}\caption{{\bf Groundstate preparation of a Hamiltonian.}
\textbf{a)} Post-select circuit for groundstate preparation in the form of the quantum phase estimation algorithm. The algorithm prepares the groundstate of a Hamiltonian onto target qubits inputted as an equal superposition of all possible bit strings. The circuit consists of a synthesised unitary $A\text{=}\exp{(\text{-}iH\tau)}$ acting on the target qubits whilst being controlled by $m$ precision ancillae. Measuring $\ket{0}^{\otimes m}$ on the ancillae indicates a successful groundstate preparation onto the target qubits. Any other measurement indicates that an excited state has been prepared and failured. Upon failure of groundstate preparation, all qubits are reset and the circuit is repeated. 
\textbf{b)} Ancillae thermalization circuit for an equivalent $O(N)$ attempts of post-selection, with $N=2$ applications of the circuit given in a). The NOR gate compiles conditions from the ancillae whilst the reset gate, $S$, redistributes the weights of the incorrectly prepared states of the target qubits onto all bit strings.
Each iteration acts on the scrambled states of the target qubits and is controlled by the output of the last NOR ancilla. A complete description for the NOR and scrambling gate can be found in Appendix~\ref{appendix:scramb and nor}.
\textbf{c)} Fidelity between the finalised target qubits and groundstate of the Hamiltonian using ancillae thermalization. We show results for different numbers of iterations. The results were obtained from IBMQ's `qasm-simulator' for $H_1 = 0 \ket{0}\bra{0} - \frac{3\pi}{2}\ket{1}\bra{1}$ and $H_2 = \sum_{i,j=1}^4 a_{ij}\ket{i}\bra{j}$. We simulate different levels of noise based upon a model of thermal relaxation between the qubits and their environment. The range of data points for the noise based simulations was restricted due to computational limitations. Additionally an approximation was made on the initialization and scrambling operations in the non-diagonal two qubit case. Further details of these experiments can be found in Appendix~\ref{appendix:scramb and nor}. }
\label{fig:Phase estimation}
\end{figure*}
Next, we apply our procedure to a groundstate preparation algorithm. 
Although more efficient state preparation algorithms exist, see \cite{Yimin, wiebe2016efficient,lin2020near}, this setting is still of interest since it reveals the role of the ancillae as an effective low-temperature bath in addition to demonstrating a FP OAA scheme for non-unitary transformations. 

The quantum phase estimation algorithm shown in Fig.~\ref{fig:Phase estimation}a), computes the eigenvalue $\theta$ satisfying $A\ket{A} = \exp{(2\pi i \theta)}\ket{A}$. Post-selecting on the precision qubit register can prepare target qubits in the groundstate of an n-qubit Hamiltonian. Ancillae thermalization achieves the same effect by tracing out ancillae qubits --- the ancillae effectively provide a low entropy reservoir into which the excess energy of the target state can be transferred.
\\
The circuit shown in Fig.~\ref{fig:Phase estimation}b) achieves the same level of accuracy as $O(N)$ for $N=2$ attempts of phase estimation with post-selection. In a similar manner to the quantum perceptron, a total overlap with the groundstate within additive error $\epsilon$ is achieved by applying $V$ to fresh ancillae and the target qubits $O(2^m\log(1/\epsilon))$ times.
After tracing out the ancillae qubits, the fidelity between the finalised mixed target state $\rho$ and the groundstate is given by,
\begin{equation}
 \bra{\lambda_G} \rho \ket{\lambda_G}) 
 = 
 \left(1 - \frac{J}{N}^{\log\left(\epsilon\right)/\log\left(J\right) + 1}\right),
\label{eq:Fidelity: Phase Estimation}
\end{equation}
where $J= (N - 1)/N$, $N = 2^n$ is the number of eigenstates and $m$ is the minimum number of precision qubits required to distinguish between all energy values without imperfections. 
Therefore the precision, i.e. the number of ancillae used in the phase estimation circuit, dictates an upper bound on the fidelity. Ancillae thermalization shows an exponential increase in fidelity as the number of iterations increase compared to the fidelity attained with the same number of attempts of the post-select circuit.
We assume that the value of the groundstate energy is known up to precision $2^{-m}$.
Additionally, a pre-processing procedure has occurred which shifts all energy values by this amount such that correctly preparing the groundstate is indicated by measuring $\ket{0}^{\otimes m}$.
\\
Results of applying the ancillae thermalization to groundstate preparation obtained from IBMQ's `qasm-simulator' with the addition of simulated noise are shown in Fig.~\ref{fig:Phase estimation}c).
The fidelity was computed between the finalised target qubits and the groundstates of the one qubit Hamiltonian $H_1 = 0 \ket{0}\bra{0} - \frac{3\pi}{2}\ket{1}\bra{1}$ and the two qubit Hamiltonian $H_2 = \sum_{i,j=1}^4 a_{ij}\ket{i}\bra{j}$ (a random set of parameters $a_{ij}$ were chosen in the latter case as described in Appendix~\ref{appendix:Numerical Simulation Parameters})  for different numbers of iterations of the ancillae thermalization circuit. As in the case of the quantum perceptron an increase in fidelity with the number of iterations reaches an upper bound when the circuit depth leads to too great an exposure to gate noise. 
\\
The fidelity is lower in the two-qubit case than predicted analytically. This is due to an approximation made on the initialization and scrambling operations on the target qubits. Furthermore, due to the inclusion of Toffoli gates in the NOR gate, the ancillae thermalization modification of post-selection is highly sensitive to noise. This is exacerbated for larger Hamiltonians due to the increase in number of gates required to act on the target qubits and slower convergence of fidelity with the number of iterations. A more detailed discussion of these effects and of the parameter values used in the simulations can be found in Appendix~\ref{appendix:Numerical Simulation Parameters}.
\\
OAA schemes alone cannot deterministically implement non-unitary transformations. However, recent developments in block-encoding \cite{berry2015hamiltonian} and quantum signal processing \cite{gilyen2019quantum} allow us to embed an approximate $n$-qubit projector in the upper-left corner of an $n+k$ qubit unitary, where $k$ is the number of ancillae needed for the encoding. Amplitude amplification is then used to deterministically implement this approximate projector onto the target qubits \cite{lin2020near}.
A further comparison of resource costs between these methods can be found in table~\ref{tab:Groundstate preparation resources} in Appendix~\ref{Appendix:Resource Scaling}.
State of the art algorithms for groundstate preparation assume the initial target qubit state has a non-trivial overlap with the groundstate.
For ease of demonstration we initialize the target state in an equal superposition of all eigenstates and construct a reset gate, i.e. the scrambling gate, which scrambles every eigenstate such that the output has an equal overlap with all other eigenstates.
In order for ancillae thermalization to have a competitive groundstate preparation algorithm, a more sophisticated initialization and reset procedure must be implemented.
\begin{figure*}[ht] 
  \hspace{-0.5cm}\begin{minipage}[b]{0.5\linewidth}\raisebox{94pt}{\textbf{(a)}}
    \centering
    \includegraphics[width = 7.0cm]{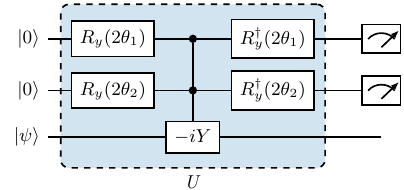}
    \vspace{4ex}
  \end{minipage}%%
  \hfill
  \begin{minipage}[b]{0.5\linewidth}\hspace{-1.35cm}\raisebox{105pt}{\textbf{(b)}}
    \centering
    \vspace{-10pt}\includegraphics[width = 8.0cm]{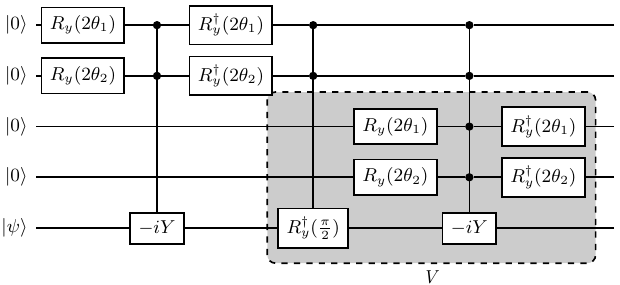} 
    \vspace{4ex}
  \end{minipage}
  \begin{minipage}[b]{0.5\linewidth}\hspace{-0.5cm}\raisebox{155pt}{\textbf{(c)}}
    \centering
    \includegraphics[width = 7.0cm]{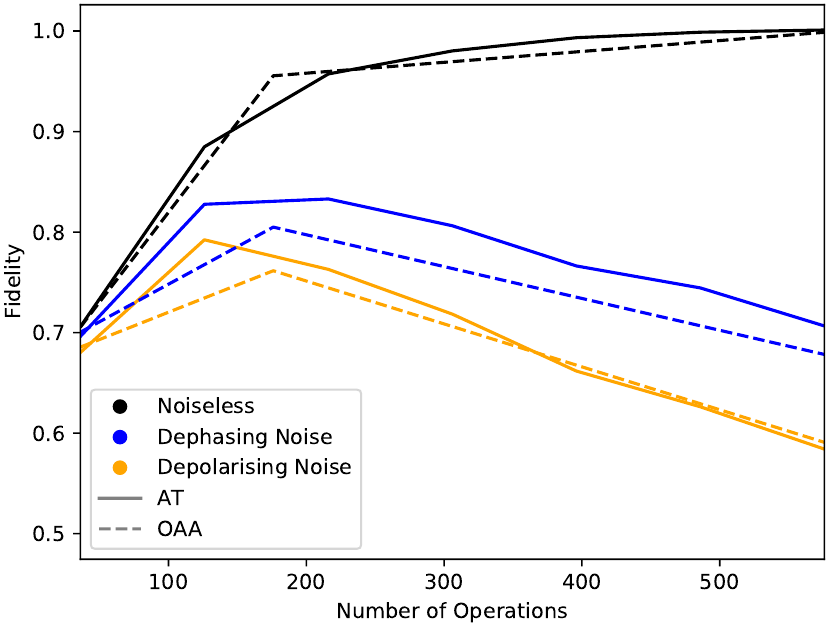} 
    \vspace{4ex}
  \end{minipage}%%
  \hfill
  \begin{minipage}[b]{0.5\linewidth}\hspace{-2.4cm}\raisebox{155pt}{\textbf{(d)}}
    \centering
    \includegraphics[width = 7.0cm]{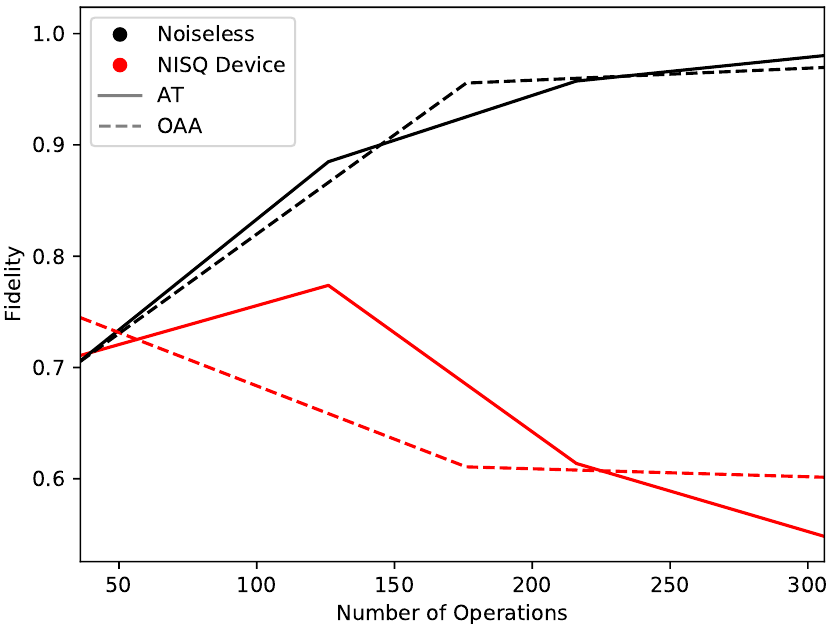} 
    \vspace{4ex}
  \end{minipage}
\caption{\textbf{Noise Robustness for the Quantum Gearbox. a)} Quantum gearbox circuit for $m=2$ ancillary qubits \cite{wiebe2013floating}. This circuit is a generalisation of the quantum perceptron found in Fig.~\ref{fig:Quantum Perceptron} with $m=1$ ancilla. Measuring $\ket{0^{\otimes m}}$ on the ancillae with probability $p\left(\theta\right) = \cos^4(\theta) + \sin^4(\theta)$ corresponds to the successful transformation of $\exp\left(-iq(\theta)Y\right)$ on the target qubit where $q(\theta) = \arctan(\tan^2(\theta))$ and $\sin(\theta) = \sin(\theta_1)...\sin(\theta_m)$. Measuring $\ket{1}$ on any ancillae corresponds to applying $R_y(-\pi/2)$ onto the target qubits, and thus can always be reset by applying $R_y(\pi/2)$. For the simulation $\theta_1 = \theta_2 = \pi/4$. \textbf{b)} Ancillae thermalization circuit for an equivalent $O(N)$ attempts of post-selection with $N=2$ applications of the circuit in a). Unitary $V$, which acts on the fresh ancillae and incorrect states of the target qubit, is conditioned by the state of all ancillae from the previous iteration. \textbf{c/d)} Fidelity between the finalised target qubit and desired state, $\exp\left(-iq(\theta)Y\right)\ket{\psi}$ using ancillae thermalization and $\pi/3$ FP OAA for different number of operations i.e. exposure to gate noise. Both dephasing and depolarising noise models were simulated in c) whilst d) is ran on `ibmq\_vigo' quantum device, in addition to a noiseless simulation.}\label{fig:at vs oaa noise} 
\end{figure*}

\section{Robustness To Noise}\label{appendix:noise}
%%%%%%%%%%%%%%%%%%%%%%%%%%
Resource costs such as the number of qubits and gate operations are a good indication of an algorithm's efficiency. On near-term quantum devices, however, an algorithm's robustness to noise is a much more practical measure. Ancillae thermalization is more robust than alternative schemes. 

{\it Demonstrating  robustness to noise:}
%%%%%%%%%%%%%%%%%%%%%%%%
Fig.~\ref{fig:at vs oaa noise} c)/d) demonstrates ancillae thermalization's robustness to noise compared with $\pi / 3$ FP OAA for the quantum gearbox --- an extension of the quantum perceptron with two ancillary qubits \cite{wiebe2013floating}.
The circuits for the quantum gearbox, ancillae thermalization and $\pi / 3$ FP OAA are given in Fig.~\ref{fig:at vs oaa noise}a, Fig.~\ref{fig:at vs oaa noise}b and Fig.~\ref{fig:FP OAA circuit} in Appendix~\ref{appendix:Numerical Simulation Parameters}.
The multi-control gates in both circuits were implemented without additional ancillae.
Specifically, the 3 qubit control gate was implemented using 6 C-NOT gates and 7 single-qubit control gates, whilst the 4 qubit control gate was implemented using a 2 single-qubit control gates and 3, 3 qubit control gates \cite{barenco1995elementary}.
Simplified multi-qubit Toffoli gates were used to further reduce operational cost\cite{maslov2016advantages}.
\\
Full state tomography was performed on the target qubit with 24576 shots for both circuits.
The fidelity was then computed between the target qubit and desired transformation, $R = e^{-iY\theta}$, where $\sin(\theta) = \sin(\theta_1)\sin(\theta_2)$. 
Note that in practice full state tomography is not required as the target qubit will be assumed to have a sufficiently large fidelity with the desired state. 
Additionally, the number of single qubit and C-NOT operations were measured per iteration of each algorithm.
\\
In addition to demonstration on IBMQ's quantum machine, simulations of both circuits were performed with depolarising and thermal relaxation noise.
The depolarising noise error parameter, $\lambda = 0.001, 0.01$, for all the single qubit and 2 qubit gates, respectively. Details of the latter noise model are given by the `high' noise level in table~\ref{table: Parameter Values} in Appendix~\ref{appendix:Numerical Simulation Parameters}.
All circuits run on IBMQ's quantum machine were compiled using the OpenQASM backend \cite{cross2017open} without additional error mitigation techniques.

{\it Origin of Noise Robustness:}
%%%%%%%%%%%%%%%%%%%%%
We believe that the robustness to noise of ancillae thermalization arises because of the intrinsic robustness of the thermalization to changes in the coupling of the system to the bath. In the special case of the quantum perceptron, some of the controls --- which are the proxy for the system-bath interaction --- can be removed entirely without any detriment to the performance. This can be seen in Fig. 2 b) where no controls are placed upon the $R_y(2 \theta)$ rotations of the fresh ancillae. We have preliminary evidence of robustness to reducing controls in other circumstances. This strongly suggests that ancillae thermalization does indeed inherit robustness to noise from independence upon the bath-system interaction. A thorough analysis will be the subject of a future work.

An additional consideration in the comparison between ancillae thermalization and $\pi/3$ FP OAA is the number of operations acting on each ancilla qubit. This number is fixed in ancillae thermalization by the depth of the post-select unitary, regardless of the finalised accuracy, exposing ancillae to less gate noise.

We expect our intuition to be applicable to a variety of circuit and quantum machine archetypes.
The fixed-point quantum search proposed by Yoder \textit{et. al} has been shown to have an exponential decrease in query complexity over $\pi / 3$ fixed-point quantum search in the regime of small initial success probabilities. Currently, it is unknown whether fixed-point quantum search for OAA has an increased robustness to gate noise compared to ancillae thermalization in this regime.
In the large initial success probability regime however, it is known that the operational costs of fixed-point quantum search coincide with $\pi / 3$ FP OAA for an equivalent finalised success probability.
Therefore although a direct comparison has not been made, we expect ancillae thermalization to have the highest overall robustness to gate noise within the large initial success probability regime.

{\it Mitigating qubit costs and control complexity}:
%%%%%%%%%%%%%%%%%%%%%%%%%%
One drawback of ancillae thermalization is the use of resource intensive control gates.
However, the same considerations that suggest robustness to gate noise also motivate ways to mitigate control costs and complexity. Inspired by the fact that a system thermalizes when only a subset of its modes are coupled to a heatbath, we have preliminary evidence that the number of control qubits and complexity of controls can be reduced by conditioning on a subset of factors of the unitary. In the case of thermalization, the coupling to the bath can be simple providing that the Hamiltonian is sufficiently scrambled. The scrambling transfers energy to the bath-coupled modes where it is dissipated. We find that it is sufficient to control simple factors of the unitary with the more complex factors playing the role of scrambling. In the context of repeat until success, such a mitigation scheme would correspond to controlling on an imperfect error flag. The result would no longer be guaranteed a success. This is less useful than its application in ancillae thermalization where even an imperfect correction of error increases the amplitude for success exponentially in time. The speedup from conditioning on a subset of operations does not change the linear-in-time scaling of the number of qubits, only the pre-factor to this scaling. Moreover, the time to reach the desired accuracy increases, reflecting the increased time to thermalise if the coupling to the bath is weakened. Nevertheless, this ability to tension these costs against one another will likely prove useful in near-term applications. 

\section{Discussion}\label{sec: discussion}
%%%%%%%%%%%%%%%%%%%%%%
The non-linearity of the classical world can be understood by the observation of a minor part of a quantum system - the unobserved part of the system acting as an environment.  
The environment can be interpreted as a heatbath extracting entropy from our system, or equivalently an entanglement bath which gradually and selectively entangles with a subset of our system.
A simple and effective model of a heatbath is to assume no back reaction so that each mode of the heatbath interacts exactly once with the system of interest. It forms such a small fraction of the overall size of the bath that the bath distribution is unaltered. At the same time, the fact that the system never interacts with this mode again means that the back reaction effects are not felt. We have used these two ideas to allow a set of ancillae qubits initialised in some low entropy state to extract entropy from our system. The free evolution of our ancillae is with a zero energy Hamiltonian - ensuring that entropy only flows from the system of interest to the ancillae and each ancilla interacts only once with the system corresponding to a no back reaction condition. 
The resulting algorithm is a type of FP OAA for unitary and non-unitary transformations, achieving non-linearity by tracing out auxiliary degrees of freedom.
Its structure is rather different from its counterparts, which require fewer qubits.
The $\pi/3$ FP OAA scheme achieves an optimal amplitude amplification through a cunning cancellation of phases.
It is a fundamental observation of statistical mechanics however that the nature of a heat bath does not determine the thermal equilibrium state (provided suitably weak coupling).
Our scheme effectively harnesses this universality to obtain a degree of robustness to gate infidelity in addition to deterministically implementing a wider class of transformations without knowledge of the target state.
Moreover it seems possible to reduce the qubit cost and control gate complexity at the expense of longer times. This gives additional freedom to operate within the NISQ constraints of qubit count and gate fidelity.
Which scheme is optimal is contingent upon the particular system to which the algorithm is applied. Still, it is gratifying that there exists a regime where a simple physically motivated scheme such as the one we present can outperform other methods.

\section{Acknowledgement}\label{sec: acknowledgement}
%%%%%%%%%%%%%%%%%%%%%%
We acknowledge support from the EPSRC: LW and FB through EP/L015854/1 , FB and AGG through EP/S005021/1, JD through EP/S021582/1  and GHB through support from the Royal Society via a University Research Fellowship, as well as funding from the European Research Council (ERC) under the European Union’s Horizon 2020 research and innovation programme (Grant Agreement No. 759063)

%\clearpage
\bibliographystyle{unsrt}
%\bibliography{apssamp.bib}

\newpage
\clearpage
\appendix
\footnotesize{
\section{Resource Scaling}\label{Appendix:Resource Scaling}
%%%%%%%%%%%%%%%%%%%%%%%%%%%%%%%

\begin{table*}[ht]
\hfill
\begin{tabular}[t]{l|c|c|c}
Groundstate Preparation& Measurements & Qubits & Gates\\
\hline
  Post-Selection \& PEA & $\displaystyle O\left(2^{n}\right)$ & $\displaystyle O\left(n +\log(1/\Delta)+\log(1/\epsilon)\right)$ & $\displaystyle \tilde{O}\left(2^{n/2}\Delta^{-1}\epsilon^{-1} \Lambda\right)$ \\
  \begin{tabular}[c]{@{}l@{}}Ancillae\\Thermalization \& PEA\end{tabular} &  $\displaystyle 0$ & $\displaystyle O\left(2^{n}\text{polylog}(1/\epsilon,1/\Delta)\right)$& 
 $\displaystyle \tilde{O}\left(2^{3n/2}\Delta^{-1}\epsilon^{-1} \Lambda\right)$ \\ \hline
 \begin{tabular}[c]{@{}l@{}}Ancillae\\Thermalization \& LCU\end{tabular} &  $\displaystyle 0$ & $\displaystyle O\left(2^{n}\text{polylog}(1/\epsilon,1/\Delta)\right)$&
 $\displaystyle O\left(2^n \Delta^{-1}\Lambda\text{polylog}(2^n,\Delta^{-1},\epsilon^{-1}) \right)$ \\
 Block Encoding \& OAA &  $\displaystyle 0$  & $\displaystyle O\left(n + k\right)$ & 
 $\displaystyle O\left(2^{n/2} \Delta^{-1} k \Lambda \log(1/\epsilon) \right)$
\end{tabular}\hfill \qquad\qquad\qquad

\caption{\textbf{Comparison of Computational Resources for Groundstate Preparation.} Computational resources for post-selection and ancillae thermalization using phase estimation (PEA) as well ancillae thermalization using linear combination of unitaries (LCU) via the results found in Ref.\cite{Yimin}. We also show the results for most state-of-the-art quantum groundstate preparation algorithm \cite{lin2020near}. This algorithm uses block-encoding in addition to amplitude amplification to approximately project the target qubits onto the groundstate, where $k$ is the number of qubits required in the block-encoding. $\tilde{O}$ is up to polylogarithmic factors. Here we assume the number of calls to $H$, $\Lambda \sim d$, where $d$ is the sparsity of the Hamiltonian i.e. maximum number of non-zero elements in each row of the Hamiltonian and $\Delta$ is the lower bound on the spectral gap. Note that we assume the input target state is given as an equal superposition of all eigenstates. Various schemes may exponentially reduce the number of ancillary qubits required for ancillae thermalization, e.g. by conditioning on only a subset of elements of a factorisation of the unitary $U$. Indeed, a heat bath does not need to couple to all elements of a system to effectively cool.}
\label{tab:Groundstate preparation resources} 
\end{table*}
\noindent
The entries in Table.~\ref{tab:General post-selection vs tracing out circuit} of the main paper show a comparison of resources required to implement the techniques discussed in this paper compared to ancillae thermalization. Here we give more detail on how these entries are obtained.

\subsection{Unitary Transformation}
%%%%%%%%%%%%%%%%%
\vspace{0.05in}
\noindent
{\it Measurements:} 
For a post-select unitary $U$ acting on $m$ ancillae and target qubits $\ket{\psi}$, up to $O(2^m)$ measurements on the ancillae are required to implement $R$ onto $\ket{\psi}$. This assumes all measurements are independent of each other. On the other hand, no measurements are required in $\pi/3$-OAA and ancillae thermalization.

\vspace{0.05in}
\noindent
{\it Number of Qubits:} 
All qubits can be reused upon failure in post-select circuits since the ancillae state collapses upon measurement.
The state $E\ket{\psi}$ is independent of $R$ and can always be reset. 
Therefore $U$ acting on $m$ ancillae and $n$-target qubits requires $n+m$ qubits in total.
Using an additional $m-1$ qubits in OAA allows for a linear scaling in $Q(S_m(\pi/3))$.
To achieve a fidelity $\bra{ \psi} R^\dagger \rho R \ket{\psi} = 1- \epsilon$ for ancillae thermalization requires $N = \log \left(  1/\epsilon \right) / \log \left(  1/(1-p_0) \right) - 1$ applications of $V$, where $\epsilon(1 - |\langle{\psi}|{{\text{R}\psi}\rangle})|^2 )^{\text{-}1} \rightarrow \epsilon$.
Insertions of $m$ new ancillae and $m-1$ NOR ancillae qubits are needed at each application.
Therefore a total $O\left(n + \log \left(  1/(1-p_0) \right)\log\left(1/\epsilon\right)m\right)$ qubits are required for ancillae thermalization.

\vspace{0.05in}
\noindent
{\it Operations:} The most important resource in the comparison of ancillae thermalization and OAA is the number of gate operations required. This is computed in the case of post-selection, OAA and ancillae thermalization as follows:
\\
{\it i. Post-selection}. We define $Q(O_{N})$ as the number of single qubit and C-NOT gates required to implement the N-qubit gate $O$.
Therefore, $U$ requires a total of $Q(U_{n+m})$ operations for $n$ target and $m$ ancillary qubits .
No further coherence time nor operations are needed, since the ancillae becomes disentangled with the target qubits once measured and the circuit can be reset. 
\\
\noindent
\textit{ii. $\pi/3$ FP OAA}.
A detailed derivation of the resource costs can be found in Ref.  \cite{guerreschi2019repeat}. We summarise the key results here.
An upper bound on the error for the k'th nested iteration is given by $(1-p)^{3^k} \le \epsilon$.
Rearranging we find,
\begin{equation}
   % k = \frac{\log\log\left(\frac{1}{\epsilon}\right) - \log\log\left(\frac{1}{(1-p)}\right)}{\log 3}.
     k = \frac{1}{\log 3} \left[ \log\log\left(\frac{1}{\epsilon}\right) - \log\log\left(\frac{1}{(1-p)}\right) \right].
   \label{eq:k}
\end{equation}
The number of operations at each iteration can be computed using the recursive relation $Q(A_{n,m}(j)) = Q(A_{n,m}(j-1)) + 2Q(S_{m}(\frac{\pi}{3}))$.
Consequentially its closed form can be written as, 
\begin{equation}\label{eq:OAA gate number}
    Q\left(A_{n,m}(k)\right) = \left(Q(U_{n+m}) + Q(S_{m}(\frac{\pi}{3}))\right)3^k -  Q(S_{m}(\frac{\pi}{3})),
\end{equation}
where the initial conditions are given in Eq.\ref{eq:initial condition one} and \ref{eq:initial condition two}.

The function $Q(S_m(\pi/3))$ is the number of operations required to implement the controlled phase shift.
Assuming access to additional ancillae, $S_m(\pi/3)$ can be constructed in a similar way to the NOR gate such that the number of operations scale linearly with $m$, for $m > 2$.
\\
Using our expression for $k$ from Eq.\ref{eq:k} and $Q(S_m(\pi/3)$, the result for the total number of operations in $\pi/3$ FP OAA follows.
\\
\textit{iii. Ancillae thermalization:}
Ancillae thermalization produces an overlap $\bra{\psi} R \rho R\ket{\psi})= 1 - (1-p)^{P+1}(1 - |\langle{\psi}|{{\text{R}\psi}\rangle})|^2 )$ between the finalised target qubits and the desired state for $P$ iterations of $V$. To achieve an overlap $1- \epsilon$, $O\left(\log(1/\epsilon)\right)$ implementations of $V$ are required.
Each $V$ consists of $Q(U_{n+m}) + Q(R_n)$ operations, of which controls need only be implemented on gates acting on the target qubits.
The result for the number of operations required in ancillae thermalization follows.

\subsection{Groundstate Preparation}
%%%%%%%%%%%%%%%%%%%
\noindent

In this paper we use groundstate preparation as a demonstration to deterministically apply a non-unitary operation onto target qubits using ancillae thermalization.
We also mention using OAA in addition to linear combination of unitaries to achieve the same task.
Here we give a comparison of resources between the two schemes.

\subsubsection{Phase Estimation}
%%%%%%%%%%%%%%%%%%%%%%%%%%
The phase estimation algorithm (PEA) implements an exact projector onto the groundstate of a Hamiltonian with $p = 1 - \epsilon$. Error $\epsilon$ is intrinsic to PEA and originates from the binary approximation of the eigenvalues.
A higher precision is required to ensure these `imperfections` do not effect the computation. We assume a pre-processing shift has occurred such that the groundstate energy value is $0$ and initialise the target state in an equal superposition of all eigenstates.
\\
\vspace{0.05in}
\noindent
{\it Measurements:} 
For an initial target state, $\ket{\psi}$, given as an equal superposition of eigenstates an average $O(p_0^{-1}) = 2^{n}$ measurements for $p_0 = 2^{-n}$ are required for groundstate preparation. No measurements are required in ancillae thermalization nor block encoding + OAA.

\vspace{0.05in}
\noindent
{\it Number of Qubits:}
Post-selection and the PEA act upon an $n$-qubit input state and $m$ ancillae qubits.
Here, the number of ancillary qubits is dependent on the required precision of the eigenvalue, $m \sim O(n/2 + \log(1 / \epsilon) + \log(1 / \Delta))$, where $\Delta$ is the lower bound on the spectral gap \cite{Yimin}.
Qubits can be reused upon failure.
Ancillae thermalization with the PEA require a total $O\left( m\log(1/\epsilon) / \log(1/(1-p_0)) \right)$ qubits, the result follows with $\log(1/(1-p_0)) \sim O(2^{n})$.

\vspace{0.05in}
\noindent
{\it Operations:}
\\
{\it i. Post-selection}. $U_{n+m}$ requires $2^m$ applications of the controlled evolutionary unitary, $A=e^{-iH}$. To implement $A$ with error $\epsilon'$ requires $Q(A) \sim O(\Lambda \text{polylog}(2^n,1/\epsilon'))$ operations where $\Lambda$ is the number of operations required to simulate $H$ \cite{low2019hamiltonian}. We assume $\Lambda \sim O(d)$, where $d$ is the sparsity of $H$ i.e. maximum number of non-zero entries in each row of $H$ \cite{low2019hamiltonian} and $Q(W_n) \sim O(1)$.
Therefore, to prepare the groundstate using the PEA requires $Q(U_{n+m}) \sim O(2^m Q(A)) = \tilde{O}(2^{n/2}\Delta^{-1}\epsilon^{-1}d)$.
\\
{\it ii. Ancillae thermalization}. The number of operations for the NOR gate scale linearly with $m$ whilst two additional operations are needed per ancillae to implement a control. We require all qubits that have been acted on to remain coherent throughout the computation and assume $Q(W_n) \sim O(1)$ or $\sim O(\text{log}(n))$ \cite{dankert2009exact}. The total number of operations is $O\left(\log(1/\epsilon) / \log(1/(1-p_0)) Q(U_{n+m})\right)$, the result follows.
\subsubsection{Linear combination of unitaries}
Linear combination of unitaries (LCU) can be used to construct a truncated Taylor series of the time-dependent evolutionary operator. This approximately projects the target qubits onto the groundstate of the Hamiltonian. The implementation of this unitary is non-deterministic, thus either OAA or ancillae thermalization can be used to amplify the probability of success. A detailed derivation of the resource costs of LCU for groundstate preparation can be found in \cite{Yimin}. We summarise the key results here.

\vspace{0.05in}
\noindent
{\it Number of Qubits:}
The number of ancillary qubits i.e. the precision required for LCU is less than PEA, requiring $m \sim O(\log(1/\Delta) + \text{loglog}(2^n/\epsilon))$ qubits to achieve the same accuracy. 
The probability of correctly implementing the groundstate projector by LCU is $p_0 \sim O(2^{-n})$, assuming $\ket{\psi}$ is in an equal superposition of eigenstates.
We ignore any qubits required for the Hamiltonian simulation. 

\vspace{0.05in}
\noindent
{\it Operations:}
LCU requires implementing a unitary, $B$ on the ancillae followed by $A_k = \exp(-iHk)$ on the target qubits, representing a segment of the Taylor series for $A$.
$B$ can be implemented with $Q(B) \sim O(\Delta^{-1}\log^{3/2}(2^n/\epsilon))$ operations.
The Hamiltonian simulation represented by $A$ is implemented with $Q(A) \sim O(\Lambda \text{polylog}(2^n,\tilde{\epsilon}^{-1})$ operations to a required accuracy $\tilde{\epsilon} = \tilde{O}(2^{-n} \Delta \epsilon)$.
Thus the number of operations needed to perform LCU, $Q(LCU) \sim O(\Delta^{-1}\Lambda\text{polylog}(2^N,\Delta^{-1},\epsilon^{-1}))$. 
Assuming $Q(W_n)$ and $Q(C\text{-LCU}) - Q(\text{LCU}) \sim O(1)$ leads to the result shown in Table~\ref{tab:Groundstate preparation resources} for ancillae thermalization.

\section{Methods}\label{appendix:scramb and nor}
%%%%%%%%%%%%%%%%%%%%%%%%%
\begin{figure*}[ht]
\hspace{-1.0cm}
\begin{minipage}[c]{0.5\linewidth}
\centering
\hspace{-8cm}\raisebox{0pt}{\textbf{(a)}}
 \\
\begin{quantikz}[row sep=0.3cm, column sep=0.2cm]
    \qw & \qw & \qw\qwbundle{m} & \qw & \qw & \gate[wires=4, nwires = {2}][0.6cm]{\text{NOR}} & \qw & \qw \\
    \\
    \qw & \qw & \qw\qwbundle{m-2} & \qw & \qw & \qw & \qw & \qw \\
    \qw & \qw & \qw & \qw & \qw & \qw & \qw & \qw
    \end{quantikz}
    =\begin{quantikz}[row sep=0.3cm, column sep=0.2cm]
    \lstick[wires = 5, label style = {rotate = 90, xshift = 0.73cm, yshift = 0.6cm}]{m precision\\ ancillae}\qw & \gate{X} & \ctrl{5} &\qw & \qw & \ \ldots\ \qw & \qw & \qw & \qw \\
    \qw & \gate{X} & \ctrl{4} &\qw &\qw & \ \ldots\ \qw & \qw & \qw & \qw\\
    \qw & \gate{X} & \qw &\qw &\ctrl{3}& \ \ldots\ \qw & \qw & \qw & \qw\\
    & \vdots & & & & \vdots\\
    \qw & \gate{X} & \qw & \qw &\qw & \ \ldots\ \qw &\ctrl{4} &\qw & \qw \\
    \lstick[wires = 4, label style = {rotate = 90, xshift = 0.73cm, yshift = 0.6cm}]{m-2 NOR \\ ancillae}\qw & \qw & \targ{}  & \qw & \ctrl{1} & \ \ldots\ \qw & \qw & \qw & \qw\\
    \qw & \qw & \qw & \qw & \targ{} & \ \ldots\ \qw & \qw & \qw & \qw\\
    & \vdots & & & & \vdots\\
    \qw & \qw & \qw & \qw &\qw & \ \ldots\ \qw& \ctrl{1} & \qw & \qw \\
    \lstick[wires = 2,label style = {rotate = 90, xshift = 0.73cm, yshift = 0.6cm}]{Last NOR \\ ancilla}\qw & \gate{X} & \qw & \qw &\qw & \ \ldots\ \qw& \targ{} & \qw & \qw
    \end{quantikz}
\end{minipage}
\hspace{2.0cm}
\begin{minipage}[c]{0.4\linewidth}
\hspace{-5.0cm}\raisebox{0pt}{\textbf{(b)}}
\\
\includegraphics[width = 9.0cm]{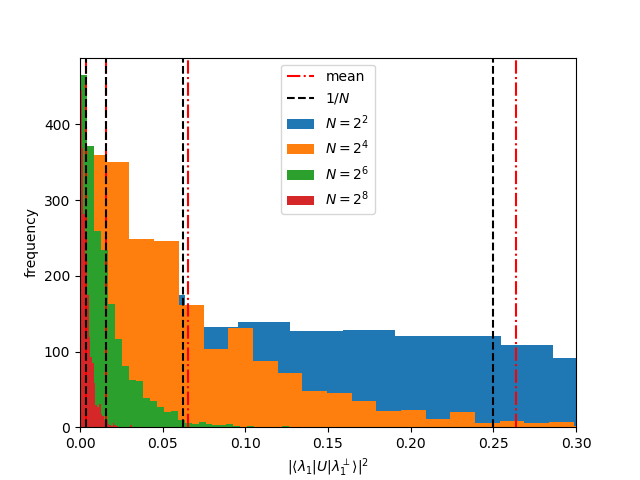}
\end{minipage}
\caption{\textbf{Operations required in ancillae thermalization for groundstate preparation.}
        \textbf{a)} The NOR gate quantum circuit used to compile the precision ancillae conditions into a single qubit. 
        The gate acts on all ancillae from the iteration and $m-1$ fresh NOR ancillae.
        The output state of the last NOR ancilla is $\ket{0}$ if and only if the state of the precision ancillae is $\ket{0}^{\otimes t}$, otherwise the last NOR qubit state is $\ket{1}$.
        The scrambling gate and next iteration of $U$ are both controlled by the last NOR qubit.
        \textbf{(b)} Histogram showing the overlap between a randomly chosen state and a scrambled state which has been transformed by an arbitrary local unitary.
        We can see numerically that the unitary increases the overlap between the states to $\sim \frac{1}{N}$, becoming more accurate as the system size increases.
        This proves that the application of a scrambling gate gives the desired result of redistributing the probability weights of each state equally.}
\label{fig:NOR and scrambling gates}
\end{figure*}
\noindent
We provide a detailed description of the methods and results found in the main text. 

\subsection{Ancillae Coupling and Reset Gates}
%%%%%%%%%%%%%%%%%%%%%%%%%
\noindent
To construct the ancillae thermalization circuit we use the methods discussed in Section~\ref{subsec: Post Select & Tracing Out} to initially apply $U$ followed by iterations of the controlled unitary $V$.
For groundstate preparation, $V$ includes a reset gate and unitary $U$ of the phase estimation circuit acting on excited states in the wavefunction of the target qubits.
It is constructed as follows:
Firstly, we only apply $V$ if the ancillae from the previous iteration are not in the state $\ket{0}^{\otimes t}$.
This state corresponds to the preparation of the target qubits in the groundstate.
The condition is checked through use of a NOT-OR (NOR) gate.
This logic gate acts upon all ancillae from the previous iteration and an additional $m-1$ NOR ancillae qubits.
The result of whether the previous ancillae correspond to the preparation of the groundstate is outputted onto the last NOR ancilla.
Secondly, a reset gate consisting of a scrambling operation acts upon the target qubits conditioned by the last NOR ancilla. The purpose of the scrambling operation is to redistribute the probability of an eigenstate to all other eigenstates equally. A full description of the NOR and scrambling gate can be found in appendices~\ref{sec:NOR} and ~\ref{sec:scram} respectively.
Finally, another application of the phase estimation unitary $U$ - controlled by the last NOR ancillary quibit acts upon the target qubits and a batch of $m$ fresh ancilae. 

\subsection{NOR Gate}\label{sec:NOR}
%%%%%%%%%%%%%%%%%%%%
\noindent
The quantum NOT-OR (NOR) gate shown in Fig.~\ref{fig:NOR and scrambling gates} is an quantum logic gate.
Its purpose is to compile the controls on all excited states represented by the ancillae onto a single qubit.
The gate acts on the $m$ precision ancillae from each iteration in addition to $m-1$ NOR ancillae in initial state $\ket{0}^{\otimes m-1}$.
The state of the last NOR ancilla is $|0\rangle$ if and only if the precision ancillae are in $|0\rangle^{\otimes m}$, otherwise the NOR qubit is in state $|1\rangle$.
The scrambling gate, $W$ and $U$ in the next iteration of $V$ are both controlled by the last NOR ancilla.
If the groundstate energy $E_G \ne 0$, then a pre-processing procedure needs to be implemented to shift all the energies by a constant such that $|0\rangle^{\otimes m}$ corresponds to the preparation of the groundstate of an arbitrary Hamiltonian on the target qubits.
\subsection{Scrambling Gate}\label{sec:scram}
\noindent
The purpose of the scrambling gate is to redistribute the weight of each eigenstate equally amongst all other eigenstates on the target qubits.
We assume that an equal distribution of eigenstates corresponds to a maximally mixed set of bit strings.
This approximation is discussed further in this section below. 
Fig.~\ref{fig:NOR and scrambling gates}b shows that for a randomly chosen eigenstate of $U$, applications of arbitrary local unitaries will increase its overlap with another eigenstate.
As $N$ increases, the mean overlap between the transformed eigenstate and a perpendicular state converges to $1/N$.
To numerically prove this result, $N$-qubit state $\ket{\lambda_1}$ and its perpendicular state $\ket{\lambda_1^{\perp}}$ were chosen uniformly with the Haar measure.
The overlap between $\ket{\lambda_1^{\perp}}$ and $\ket{\lambda_1}$ acted on by local Hadamard gates i.e. $H^{\otimes N}$ was computed and the process was repeated 1000 times.
The simulation was also repeated with local $X$ gates which gave the same result.
\\
\textit{Scrambling Approximation:}
The eigenstates of a diagonal unitary correspond to single bit strings.
Consequently a maximally mixed state of bit strings corresponds to a equal distribution of eigenstates, where application of a Hadamard gate on a bit string scrambles the state entirely.
This is not true for non-diagonal unitaries, where eigenstates correspond to a linear combination of bit strings. 
However, as $N$ increases so does the number of eigenstates which have an average overlap $1/N$ with each bit string. 
Therefore the assumption that a maximally mixed set of bit strings can approximate an equal distribution of eigenstates for an arbitrary unitary becomes more accurate as the number of target qubits increase.

\subsection{Fidelity Calculations}
%%%%%%%%%%%%%%%%%
\noindent
In this section we derive the fidelity between the finalised target qubits from the tracing out method and the desired state for the Quantum Perception and groundstate preparation algorithms. 
\subsubsection{Quantum Perceptron}
\noindent
Applying the Quantum Perceptron post-select unitary $U$ onto the state $\ket{0}\ket{\psi}$ produces,
\begin{equation}
    \ket{\phi_0} = \sqrt{p(\theta)}\ket{0}\ket{\psi'} + \sqrt{1-p(\theta)}\ket{1}E\ket{\psi},
\end{equation}
where $\ket{\psi'} = R(\theta)\ket{\psi}$.
The reset gate $W$ transforms $E\ket{\psi} \rightarrow \ket{\psi}$ and a new ancilla is inserted.
$U$ conditioned by the state of the previous ancilla, acts on the new ancilla and target qubits to give,
\begin{align}
    \ket{\phi_1} &= \left(\sqrt{p(\theta)}\ket{0} + \sqrt{1-p(\theta)}\sqrt{p(\theta)}\ket{1}\right)\ket{0}\ket{\psi'} \\ 
    &+ (1-p(\theta))\ket{1}\ket{1}\ket{\psi} \nonumber. 
\end{align}
Resetting, inserting new ancillae and applying $U$ conditioned by the previous iteration's ancilla $P$ times leads to the state,
\begin{equation}
    \ket{\phi_P} = \left(\sum_{k=0}^{P-1}\sqrt{p(\theta)(1-p(\theta))^k}\ket{k}\right)\ket{\psi'} + (1-p(\theta))^{\frac{P}{2}}\ket{P}\ket{\psi},
\end{equation}
where $\ket{k} = \ket{1}^{\otimes k}\ket{0}^{\otimes (P-k)}$ and $\ket{P} = \ket{1}^{\otimes P}$.
The density matrix after $P$ steps is given by,
\begin{equation}
    \rho^P = \ket{A}\ket{\psi'}\bra{\psi'}\bra{A} + \ket{B}\ket{\psi}\bra{\psi}\bra{B} + ... \ ,
\end{equation}
where $\ket{A} = \sum_{k=0}^{P-1}\sqrt{p(\theta)(1-p(\theta))^k}\ket{k}$ and $\ket{B} = (1-p(\theta))^{\frac{P}{2}}\ket{P}$.
Using $\langle k' \ket{k} = 0$ for all $k \ne k'$, a partial trace is performed on the ancillae to obtain,
\begin{equation}
    \rho^P_{\text{target}} = \left(1 - (1 - p(\theta) )^P \right) \ket{\psi'}\bra{\psi'} + \left(1 - p(\theta) \right)^P \ket{\psi}\bra{\psi}.
\end{equation}
Using the equation above, the fidelity between the finalised target qubits and desired state $F = \text{Tr}(\rho^P_{\text{target}}\ket{\psi '}\bra{\psi '})$ can be written as,
\begin{equation}
    F(\rho^P_{\text{target}},\ket{\psi'}) = 1 - (1-p)^P(1 - |\bra{\psi}\psi'\rangle|^2).
\end{equation}

As the quantum perceptron is an example of a single ancilla repeat-until-success circuit, the result can easily be generalised to the $m$ ancillae case.

\subsubsection{Groundstate Preparation}
%%%%%%%%%%%%%%%%%%%%%
\noindent
To prepare the target qubits in the groundstate of a specified Hamiltonian we utilise quantum phase estimation.
This algorithm computes $0 \le\theta \le 1$ which satisfies, $A \ket{\lambda} = e^{2\pi i \theta}\ket{\lambda}$ up to a finite precision for $m$ ancillae qubits in the first register.
In other words it computes the binary value $\tilde{\theta} = 0.\theta_1\theta_2\ ...\theta_m$ with $\theta_i\in\{0,1\}$.
The unitary $A$ can always be constructed from hermitian matrix $H$ such that $A \ket{\lambda} = e^{-iE\tau}\ket{\lambda} \ \Rightarrow \ E = -\frac{2\pi \theta}{\tau}$ where $E$ is the energy corresponding to eigenstate $\ket{\lambda}$.
Note that in all of our experiments $\tau = 1$.
\\
\\
The $n$-target qubits are initialised to an equal superposition $\ket{\psi_n} = \frac{1}{\sqrt{N}}\sum_{i=1}^{N}\ket{\lambda_i}$ of all eigenstates.
Implementing the phase estimation unitary $U$ onto $\ket{0^{\otimes m}}\ket{\psi_n}$ gives,
\begin{equation}
    \ket{\phi_{1}} = \frac{\ket{\theta_G}}{\sqrt{N}}\sum_{i=1}^{N^*}\ket{\lambda_i} + \sum_{i=N^*}^{N}\frac{\ket{\theta_i}\ket{\lambda_i}}{\sqrt{N}},
\end{equation}
The false positive in the prepared groundstate originates from the finite precision on the eigenvalues. 
A scrambling operation $S$ is performed on the incorrectly prepared eigenstates by placing a condition on the ancillae.
This operation produces an overlap $\bra{\lambda_i}S\lambda \rangle \sim \frac{1}{N}$ for all $i=1,...,N$.
Details of $S$ and the conditioning on the ancillae can be found in the section \ref{appendix:scramb and nor}.
A batch of $m$ new ancillae are inserted.
$U$ acts on the new ancillae and target qubits conditioned by the previous ancillae to transform the state as,
\begin{equation}
\ket{\phi_2} = \Bigg(\frac{\ket{\theta_G}\ket{\theta_G}}{\sqrt{N}} + \frac{\ket{\theta_G}}{N}\ket{k}\Bigg)\sum_{j=1}^{N^*}\ket{\lambda_j} + \frac{\ket{k}\ket{k}\ket{\psi_n}}{N},
\end{equation}
where $\ket{k} = \sum\limits_{i = N^* + 1}^{N}\ket{\theta_i}$ and a conditioned $S$ has been applied to the target qubits.
\\
After $P$ iterations of inserting ancillae, applying $U$ and $S$, the state is given by,
\begin{align}
    \ket{\phi_{P1}} &=\left(\sum_{i=0}^{P-1}\frac{1}{N^{\frac{i+1}{2}}}\ket{\theta_G}^{\otimes (P-i)}\ket{k}^{\otimes i}\right)\sum_{j=1}^{N^*}\ket{\lambda_j} \\
    &+ \frac{1}{N^{\frac{P}{2}}}\ket{k}^{\otimes P}\ket{\psi_N}.\nonumber
\end{align}
By expanding $\ket{\psi_N}$ it can be shown that scrambling the state increases the overlap with the groundstate, 
\begin{align}
        \ket{\phi_{P2}} &=\bigg(\sum_{i=0}^{P-1}\frac{1}{N^{\frac{i+1}{2}}}\ket{\theta_G}^{\otimes (P-i)}\ket{k}^{\otimes i} \\
        &+ \frac{1}{N^{\frac{P+1}{2}}}\ket{k}^{\otimes P} \bigg)\Tilde{\ket{\lambda_G}} 
        + \frac{1}{N^{\frac{P+1}{2}}}\ket{k}^{\otimes P}\ket{\lambda_{E}},
\end{align}
where $\Tilde{\ket{\lambda_G}} = \sum_{j=1}^{N^*}\ket{\lambda_j}$ and $\ket{\lambda_{E}} = \sum_{j=N^* + 1}^{N}\ket{\lambda_j}$.
\\
\noindent
The density matrix after $P$ steps is given by,
\begin{equation}
    \rho^P = \ket{A}\Tilde{\ket{\lambda_G}}\Tilde{\bra{\lambda_G}}\bra{A} + \ket{B}\ket{\lambda_{E}}\bra{\lambda_{E}}\bra{B} + ... \ ,
\end{equation}
where, 
\begin{align}
    \ket{A} &= \left(\sum_{i=0}^{P-1}\frac{1}{N^{\frac{i+1}{2}}}\ket{\theta_G}^{\otimes (P-i)}\ket{k}^{\otimes i} + \frac{1}{N^{\frac{P+1}{2}}}\ket{k}^{\otimes P} \right), \nonumber\\
    \text{and} \nonumber\\
    \ket{B} &= \frac{1}{N^{\frac{P+1}{2}}}\ket{k}^{\otimes P} \nonumber.
\end{align}
Performing a partial trace on the ancillae the density matrix of the target qubits is given by,
\begin{align}
    \rho^P_{\text{target}} &= \frac{1}{N^*}\left(1 - \left(\frac{N-N^*}{N}\right)^{P+1}\right)\Tilde{\ket{\lambda_G}}\Tilde{\bra{\lambda_G}} \\
    &+ \frac{(N-N^*)^P}{N^{P+1}}\ket{\lambda_{E}}\bra{\lambda_{E}} + ... \ \nonumber.
\end{align}
Using the equation above, the fidelity between the finalised target qubits and groundstate can be written as,
\begin{equation}
    F(\rho_{\text{target}}^P,\ket{\lambda_G}) = \frac{1}{N^*}\left(1-\frac{\left(N-N^*\right)^P}{N^{P+1}}\right).
\end{equation}
The fidelity is bound by the distinguishability of the eigenstates. In the main paper we choose $m$ such that $N^* = 1$. 
\section{Numerical Simulation Parameters}\label{appendix:Numerical Simulation Parameters}
\begin{figure*}[ht]
\hspace{-1.0cm}
\begin{minipage}[c]{\linewidth}
\centering
\begin{quantikz}[row sep=0.2cm, column sep=0.1cm, ampersand replacement=\&]
            \lstick{\ket{0}} \& \gate{R_y(2\theta_1)} \& \ctrl{2} \& \gate{R_y^{\dagger}(2\theta_1)} \& \qw\& \qw\gategroup[2,steps=2,style={dashed,
    rounded corners,fill=gray!40, inner xsep=2pt},
    background,label style={label position=below,anchor=
    north,yshift=-0.5cm}]{$S_m(\pi/3)$} \& \gate{\left(\begin{array}{cc} \
                e^{-i\frac{\pi}{3}} & 0 \\ 0 & 1 \end{array}\
                \right)} \& \qw \& \gate{R_y(2\theta_1)} \& \ctrl{2} \& \gate{R_y^{\dagger}(2\theta_1)} \& \qw\& \qw \& \gate{\left(\begin{array}{cc} \
                e^{-i\frac{\pi}{3}} & 0 \\ 0 & 1 \end{array}\
                \right)} \& \qw \& \gate{R_y(2\theta_1)} \& \ctrl{2} \& \gate{R_y^{\dagger}(2\theta_1)} \& \qw\\
            \lstick{\ket{0}} \& \gate{R_y(2\theta_2)} \& \ctrl{1} \& \gate{R_y^{\dagger}(2\theta_2)} \& \qw\& \qw \& \ctrl{-1} \& \qw \& \gate{R_y(2\theta_2)} \& \ctrl{1} \& \gate{R_y^{\dagger}(2\theta_2)} \& \qw\& \qw \& \ctrl{-1} \& \qw \& \gate{R_y(2\theta_2)} \& \ctrl{1} \& \gate{R_y^{\dagger}(2\theta_2)} \& \qw\&\\
    \lstick{\ket{\psi}} \& \qw\& \gate{-iY}\& \qw \& \qw \& \qw \& \qw \& \qw \& \qw\& \gate{iY}\& \qw \& \qw \& \qw \& \qw \& \qw \& \qw \& \gate{-iY}\& \qw \& \qw
\end{quantikz}
\end{minipage}
\caption{\textbf{$\pi/3$ FP OAA Circuit for the Quantum Gearbox.} First iteration of $\pi/3$ FP OAA for an equivalent $O(N)$ attemps of post-selection with $N=3$ applications of the circuit in Fig.~\ref{fig:at vs oaa noise}a. The circuits consists of repetitions of the post-select unitary $U$ and controlled phase gate $S_m(\pi/3)$ for $m=2$, which performs a phase shift on the $\ket{0}$ state of the top ancillary qubit. The quantum gearbox with $\pi/3$ FP OAA was simulated with depolarizing and thermal relaxation noise in addition to being ran on IBMQ's quantum device. Fidelity between the finalised target and desired state was computed for different numbers of nested iterations and compared against ancillae thermalization. In all 3 noise models it is shown in Figs.~\ref{fig:at vs oaa noise} (c) and \ref{fig:at vs oaa noise}(d) that ancillae thermalization has an increased robustness to gate noise as a function of circuit depth.}
\label{fig:FP OAA circuit}
\end{figure*}
\noindent
Here we provide the parameters used in the numerical simulations shown in the main text.
\subsection{Qiskit Code}
\noindent
We construct the example circuits discussed in this paper using the python qiskit API \cite{Qiskit}. This allows the decomposition of our circuits into the universal gate set consisting of arbitrary single qubit rotations and C-NOT gates.
We use a total of 8192 shots for each data point in both examples and chose not to display error bars since they are statistically negligible.
Fig.~\ref{fig:FP OAA circuit} shows the $\pi / 3$ FP OAA circuit for the Quantum Gearbox, implemented using qiskit.
The code used for the experiments in this paper can be found at \cite{codewright}.
\subsection{Hamiltonians}
The Hamiltonians used in the experiments are given by,
\begin{align}
    H_1 &= \begin{pmatrix} 
            0 & 0 \\
            0 & -\frac{3\pi}{2} 
            \end{pmatrix},
            \nonumber
    \\
    H_2 &= \begin{pmatrix} 
         -0.08609 & -0.22467 & -0.41822 & -0.10511\\
         -0.22467 & -1.40667 & -0.16506 & -0.67003 \\
         -0.41822 & -0.16506 & -3.06202 &  0.09996 \\
         -0.10511 & -0.67003 & 0.09996 & 1.41319
        \end{pmatrix}.
\end{align}
$H_2$ was chosen such that $A = e^{-iH_2} = V D V^{\dagger}$ where $D = \sum_{k=0}^3 e^{\frac{k \pi i }{2}}\ket{k}\bra{k}$.
$V$ contains the eigenvectors of $A$ and has the form, $V = \mathcal{I} + \epsilon B$ where the elements $B_{ij} \sim \mathcal{N}(0,0.5)$ and perturbation $\epsilon = 0.5$.
Orthonormalization of $V$ is ensured by the Gram-Schmidt process.
The motivation behind $H_2$ comes from the accurate approximation on the scrambling gate discussed in appendix~\ref{sec:scram}.
As the number of target qubits increase, the space of applicable Hamiltonians increases and this approximation becomes more accurate.
 
\subsection{Simulated Noise}
%%%%%%%%%%%%%%
\noindent
The simulated noises for groundstate preparation represents the thermal relaxation between each qubit and their environment. 
This was parameterised by the thermal relaxation time $T_1$, the dephasing constant $T_2$ and the implementation time of each gate.
The thermal relaxation noise model provided by Qiskit was used in the groundstate preparation experiment of $H_2$.
This model is parameterised by the thermal relaxation time $T_1$, dephasing constant $T_2$ and implementation time of: CC-A, C-A, C-NOT and single qubit gates.
Table~\ref{table: Parameter Values} shows the range of parameter values used in the groundstate preparation experiments with different levels of noise.
The noise was computed by decomposing each gate into C-NOT and single qubit gates where the noises are given explicitly.
The gates C-A and CC-A were an exception to this decomposition and custom gate noises were computed respectively using the values below. 
$T_1$ and $T_2$ were sampled for each qubit from a normal distribution with means $\mu_1$ and $\mu_2$ respectively and shared variance $\sigma$.
\begin{table}[h]
\begin{tabular}{c|cccccc|ccc}
    Noise Level         & \multicolumn{6}{c|}{Implementation Time (ns)}                                       & \multicolumn{3}{c}{\begin{tabular}[c]{@{}c@{}}Statistical\\  Params.  $(\mu s)$\end{tabular}} \\
             & $U_1$ & $U_2$ & $U_3$) & C-NOT & C-A  & CC-A & $\mu_1$                        & $\mu_2$                        & $\sigma$                       \\ \hline
Low     & 0             & 50                 & 100                      & 300   & 1600 & 3000 & 1800                           & 2000                           & 10                             \\
Medium & 0             & 50                 & 100                      & 300   & 1600 & 3000 & 180                            & 200                            & 10                             \\
High    & 0             & 50                 & 100                      & 300   & 1600 & 3000 & 50                             & 70                             & 10                            
\end{tabular}
\caption{Table of parameter values used in the simulation of thermal relaxation noise for the two qubit groundstate preparation experiment.}
\label{table: Parameter Values}
\end{table}
}
\section{Background}\label{appendix:Background}
\noindent
Here we provide a brief explanation of the other techniques mentioned in the main text.
\subsection{Classical perceptron}
\noindent
The classical perceptron whose quantum analogue is discussed in the paper consists of two parts:
 The first part takes $n$-inputs $x_1,...,x_n$ and performs linear regression with synaptic weights $w_1,...,w_n$ plus a bias $b$. This computes the input signal to the perceptron $\theta = x_1w_1 + x_2w_2 + ... + b$.
 The second part maps $\theta$ onto the activation function $a(\theta) \in [0,1]$. This is known as the state of a perceptron and is used either as an input for a next perceptron or an output for a neural network. Within the quantum perceptron the latter of the two processes is represented by an angle of rotation upon a target qubit as a function of $\theta$. The challenge is to overcome the innate linearity of quantum dynamics to find a realisation of this non-linear function.
\subsection{Oblivious Amplitude Amplification}
\noindent
Oblivious Amplitude Amplification (OAA) replaces post-selection with tracing out ancillary qubits to guarantee a specified unitary transformation, without knowledge of the target state \cite{berry2014exponential}. The allocation of resources for OAA differ to that of our proposed method. We focus upon an implementation that monotonically decreases the error of implementing the specified unitary transformation in the regime of large initial success probabilities, $\frac{\pi}{3}$ {\it FP OAA}\cite{grover2005fixed}.
%1808.02900

Repeating Eq.(1) of the main text here for clarity, we seek a transformation $U$ that achieves a desired unitary transformation $R$ of a target set of qubits with some probability $p_0$:
\begin{equation}
 U \ket{0}^{\otimes m} \ket{\psi} 
 = 
 \sqrt{p_0} \ket{0}^{\otimes m} R\ket{\psi} + \sum_{k=1}^{2^m - 1} \sqrt{p_k} \ket{k} E_{k}\ket{\psi}.
\end{equation}
In essence, $\frac{\pi}{3}$ FP OAA `boosts' the final success probability from $p_0= 1- \epsilon$ to $1-\epsilon^3$ using the equality $(1- e^{i \pi /3})=e^{-i \pi/3}$. This is done by replacing $U$ with $A_1$ given by
\begin{eqnarray}
A_0 &=&U \label{eq:initial condition one},\\
A_k &=& -A_{k-1}S(\pi/3)A_{k-1}^{\dagger}S(\pi/3)A_{k-1} \label{eq:initial condition two},
\end{eqnarray}
where $S(\pi/3)=  \mathbb{I}^m - (1-e^{i\pi/3})| 0^m \rangle \langle 0^m |$ is a controlled $\pi/3$ phase shift applied to the ancillary qubits. $A_k$ concatenates this procedure $k$ times to obtain a final success probability 
$p_{\text{final}}= 1 - \epsilon^{3^k}$. Each recursion increases $p_{\text{final}}$ super-exponentially at the cost of an exponential number of operators.
The larger number of gate operations acting on each ancillary qubit in $\pi/3$-FP OAA, as discussed in Sec.~\ref{appendix:noise}, may lead to a reduced robustness to noise when compared with ancillae thermalization.
\nocite{*}
\end{document}